\documentstyle [11pt,epsf]{article}  
\title{ The Wave Function in Quantum Mechanics \\[1.4cm] }        
\author{\it  Kiyoung Kim\/ \\
        \it Department of Physics, University of Utah, SLC, UT 84112 USA\/ \\ } 
   
\begin{document}
\pagenumbering{arabic}
%
\date{ }
\maketitle
\centerline{\\[0.5cm]} 
%
\begin{abstract}
\noindent 
\noindent    
Through a {\it new interpretation\/} of Special Theory of Relativity and with 
a model given for physical space,  we can find a way to understand  
the basic principles of Quantum Mechanics consistently from Classical Theory. 
 It is supposed that natural phenomena have a connection with {\it intangible 
reality\/} which cannot be measured directly.  
Futhermore,  the intangible reality is  supposed as vacuum particles -- 
vacuum electrons are chosen as a model, each of which has energy  $E {\sim} 
- m_e {c^2}$ in Dirac sea.  In addition,  {\it 4-dimensional complex space\/} 
is introduced, in which each dimension has an internal complex space.  
\end{abstract}
\centerline{\\[1.0cm]} 
{\bf Key Words : \/} intangible reality -- vacuum particles. \\
\makebox[2.7cm]{}      4-dimensional complex space. \\ 
\makebox[2.7cm]{}      internal complex space. \\
\makebox[2.7cm]{} \\
\makebox[2.7cm]{}
\section{Introduction} 
Although we don't have to review whole philosophy behind physics, at least  
we need to mention the philosophical attitude on which this article is based.  
Since $19^{th}$ century {\it positivism\/} or {\it empiricism\/} has been 
a concrete, non-compatible philosophical background in natural science; 
however,  \underline{in the view of {\it realism\/}}   
Quantum philosophy -- the Berkeley-Copenhagen interpretation -- seems running 
to an extreme case of the philosophy.    
\par \medskip
On the other hand, the {\it positivism\/} -- mainly based on phenomenological 
facts(materialism)   --  is the most clear way in discussion of natural 
science including phisics.   But, its limitation also appears in discussion of 
{\it fundamental physics\/}
\cite{BRODY}\cite{JSB}\cite{paul-1}\cite{DAVID}\cite{BRYCE}.     
\par \medskip    
To discuss the {\it realism\/} 
-- based on truth(idealism) not on materialism -- there is a good example, 
that is,  {\it Allegory of the Cave\/}\cite{PLATO}  by Socrates 
and written by Plato -- philosophers in ancient Greek.   
\begin{itemize}
\item[]
The cave represents the sensible  world, the prisoners are nonphilosophers, 
and their fire represents the sun.  The upper world outside the cave 
represents the intelligible realm, the released prisoners are philosophers, 
and the sun represents the good.\cite{RAY}\cite{paul-2}
\end{itemize}
In the parable of the cave, the dwellers  cannot recognize the truth 
-- puppets themselves --  but can perceive only shadows of the puppets.  
This is a metaphor which means the limitation  of phenomenological facts 
in presenting  truth.   For instance, the dwellers -- as positivists -- 
can consider the shadows themselves as truth.   Then, how often they might 
have to ask God to explain their unsolved problems?   
\par \medskip 
In this respect, {\it ontological\/} review of fundamental physics is 
necessary, then some {\it epistemological\/} interpretation should be 
followed.  The purpose of this article is not to make Quantum Mechanics little 
but to understand it comprehensively with Special Theory of Relativity because 
real problems do not go away but merely change their appearance.

\paragraph{}             
In Quantum Mechanics,  what is the physical entity represented by the    
{\it wave function\/}?  
and how can we understand {\it one particle double slits experiment\/}?   
Should we accept the wave function itself as a physical entity?  
If that is true, how can we understand {\it wave function collapse\/} from  
finite probability to absolute one in the measurement of physical quantity?    
Even though Quantum Mechanics is so useful to get some practical results,  
we might have overlooked something more fundamental. 
\medskip
\par 
In Special Theory of Relativity, how can we understand  
why the speed of light is constant to all inertial observers and where 
the increased energy(or mass) is comming from?   
However, according to the theory itself, the mass increase depends on only 
relativistic motion between two inertial frames regardless of their 
interaction.
\medskip 
\par
With a model for physical space and light propagation,   
we find a way to understand the physical entity of the {\it wave function\/}  
in Quantum Mechanics.  It is assumed that negative-energy-mass particles 
in Dirac sea be considered in physical interactions, 
and that physical space consist of 4-dimensional complex space, in another 
words, each dimension has real and imaginary parts.
\medskip 
\par 
In Section 2, without any Quantum Mechanical concept a universal constant is 
searched in Special Theory of Relativity itself. In Section 3, 4-dimensional 
complex space is suggested for physical space, and the relations between the 
real and imaginary space are assumed.  With a model suggested in Section 3.1,  
light propagation mechanism and the physical meaning of {\it photon\/} 
are studied.  In Section 4, the interpretation of inertial mass and physical 
momentum is searched, and then   the physical entity of {\it wave function\/} 
in Quantum Mechanics  and the relation with physical momentum are searched.  
In Section 5,  Fundamental questions -- 
wave function collapse,  one-particle-double-slits experiment, and 
photoelectric effect -- are discussed.  Finally,  conclusion follows 
at the end.     
\section{A universal constant for light} 
\paragraph{} 
In Special Relativity, we know that the wave vector(${k_o}$, ${\vec k} $) 
of light  
and the momentum ($p_o$, ${\vec p} $) are Lorentz 4-vectors with null lengths 
because of {\it phase invariance\/} in the wave vector and 
{\it energy-momentum conservation\/}, respectively.         
In proper Lorentz transformation, we can find that the ratio of $k_{\mu}$ to 
$p_{\mu}$ in each componemt is Lorentz scalar,  which is independent of  
Lorentz transformation as long as the momentum ${\vec p}$ is parallel to 
the wave vector ${\vec k}$. Although we already know that the constant is 
Plank's  constant($h$), with one assumption  we can confirm the universal 
constant from Special Relativity itself without using any Quantum Mechanical 
concept.   
\par  
Now let us think a light wave motion(electromagnetic wave) in free space and  
assume that the momentum($p_{\mu}$) and the wave vector($k_{\mu}$) are 
unique to describe the wave motion at least in the energy and momentum.   
Suppose that those two vectors are parallel to each other in their space 
components. If we can assume that 
${\vec p} = {\alpha}{\vec k}$ (${\alpha}$ : constant), it is so trivial; 
however, we cannot be sure if the ${\alpha}$ is a function of ${k_o}$, ${p_o}$ 
or both in the inertial frame. 
\paragraph{}
\begin{figure} 
\begin{center} 
\leavevmode 
\hbox{%
\epsfxsize=4.5in 
\epsffile{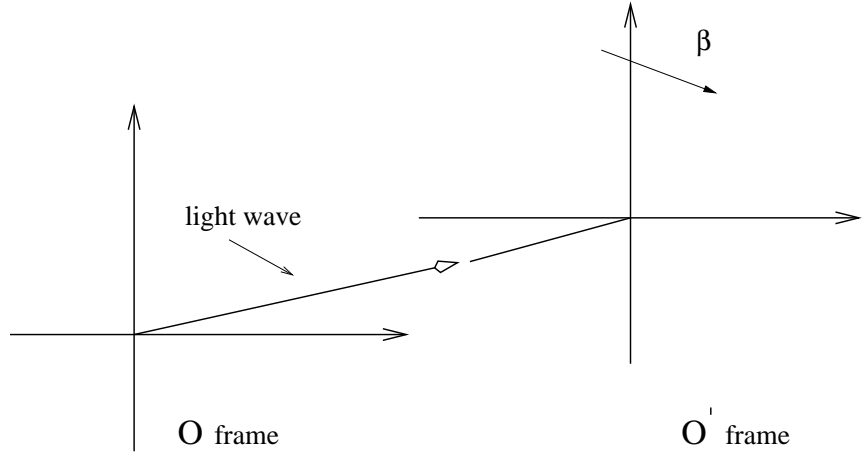}} 
\end{center} 
\caption{Lorentz transformation with arbitrary ${\vec {\beta}}$} 
\label{fig1}  
\end{figure} 
Fig.(\ref{fig1}) represents a Lorentz transformation for an arbitrary 
velocity  ${\vec{\beta}}$.  Let us say, ${k_{\parallel}^{\prime}}$ is 
parallel component and ${k_{\bot}^{\prime}}$ is perpendicular component 
to the direction of ${\vec {\beta}}$ in ${O^\prime}$-frame.     
The relations in the Lorentz transformation are  
$$     
{k_{\parallel}^{\prime}} = k {\gamma}(\cos{\theta} -{\beta}),
\makebox[0.5cm]{}  
({\gamma} = {1\over{\sqrt{1 - {\beta}^2}}}, \makebox[4mm]{}{k = |{\vec k}| = 
k_o = {\omega \over c}}) \makebox[0.5cm]{}    
$$
$$
{k_{\bot}^{\prime}} = k_{\bot}, \makebox[8.6cm]{}       
$$
$$ 
{k_o^{\prime}} = k_o {\gamma} ( 1 - {\beta} {\cos{\theta}}), \makebox[4cm]{}
\makebox[2.5cm]{}  
$$ 
in which  ${\theta}$ is the direction of cosine angle between  ${\vec k}$ 
and ${\vec {\beta}}$ in O-frame.   Likewise, the momentum $p_{\mu}$ has to 
be the same functional form as the wave vector $k_{\mu}$,   
that is
$$     
{p_{\parallel}^{\prime}} = p {\gamma}(\cos{\theta} -{\beta}),
\makebox[0.5cm]{}  
({\gamma} = {1\over{\sqrt{1 - {\beta}^2}}}, \makebox[4mm]{}{p = |{\vec p}| = 
p_o = {E \over c}}) \makebox[0.5cm]{}   
$$
$$
{p_{\bot}^{\prime}} = p_{\bot}, \makebox[8.6cm]{}       
$$
$$ 
{p_o^{\prime}} = p_o {\gamma} ( 1 - {\beta} {\cos{\theta}}). \makebox[4cm]{}
\makebox[2.5cm]{}  
$$ 
For each parallel or perpendicular component in $(p_o , \vec p)$ and 
$(k_o , \vec k)$,  the ratios of the momentum to the wave vector are   
$$  
{{p_{\parallel}^{\prime}}\over {k_{\parallel}^{\prime}}}
\makebox[2mm]{} = \makebox[2mm]{}{p \over k}  
\makebox[2mm]{} = \makebox[2mm]{}{p_{\parallel} \over k_{\parallel}} 
\makebox[2mm]{} \equiv \makebox[2mm]{} {f_{\parallel}},    
\makebox[6cm]{}  
$$ 
$$         
{{p_{\bot}^{\prime}}\over{k_{\bot}^{\prime}}} 
\makebox[2mm]{}=\makebox[2mm]{} {p_{\bot}\over k_{\bot}} 
\makebox[2mm]{} \equiv \makebox[2mm]{}{f_{\bot}}, 
\makebox[7cm]{}  
$$ 
$$    
{{p_o^{\prime}}\over {k_o^{\prime}}} 
\makebox[2mm]{}=\makebox[2mm]{} {p_o \over k_o} 
\makebox[2mm]{} \equiv \makebox[2mm]{} {f_o}. \makebox[7.4cm]{}      
$$ 
Here, we defined $f_{\parallel}$, $f_{\bot}$, and $f_o$ -- Lorentz scalars  
since the ratios are independent of $\vec{\beta}$ and $\theta$.   
In addition, we already know that $f_{\parallel}$ is same to $f_o$.   
Now let us choose a $\vec{\beta}^{\prime}$ to exchange $p_{\parallel}$ 
$\leftrightarrow$ $p_{\bot}$ and $k_{\parallel}$ $\leftrightarrow$ $k_{\bot}$ 
because the ratios are Lorentz scalars. 
Then we can conclude that $f_{\parallel} = f_{\bot} = f_o \equiv f $, that is 
$$ 
{p_{\mu} \over k_{\mu}} = f \makebox[2cm]{} \makebox[4cm]
{({$\mu$} = {\small 0,1,2,3 \/}} \makebox[2.5cm]{in any inertial frame.)} 
$$
\paragraph{}
With the fact that $p{\lambda} ({\lambda = {2 \pi \over k}})$ is a Lorentz 
scalar, let us suppose that there are two light waves with wave lengths,  
${\lambda}_1$  and ${\lambda}_2$ in O-frame(Fig.\ref{fig1}). 
If the amplitudes of the waves are {\it same\/}, then the energies($p_o$) are  
determined only by the wave lengths or the frequencies 
($k={2 \pi \over {\lambda}}$, $c = {\nu} {\lambda}$) 
because there is no difference in these two waves -- in classicsl point of 
view --   
as far as our concerning is focussed on the energy and momentum only.  
Now, let us suppose that ${{{\lambda}_1}^{\prime}}$ in $O^{\prime}$-frame 
became to --  through a continuous Lorentz transformation -- 
equal to the ${{\lambda}_2}$ in O-frame, 
then ${E_1}^{\prime}$ is equal to ${E_2}$ 
because we assumed the same amplitudes.  Therefore, $p {\lambda}$ is not 
only Lorentz scalar but also a universal constant.  
\medskip  
\par        
In above argument,  we assumed same amplitudes for the light waves and thus  
confirmed a universal constant without quantum mechanical concept. But still 
we don't know if there is a fundamental energy unit in $ E = pc $ or not.  
\section{Vacuum Particles and Physical Space}    
\subsection{Physical Vacuum} 
\paragraph{}   
From Dirac's hole theory\cite{PAM}\cite{JDB-1},   
we already know the existence of negative energy mass particles in vacuum. 
Let us call those particles {\it vacuum particles\/}.  
In vacuum, all possible states are occupied with fermions of each kind, 
and an antiparticle, in real world, is a hole in vacuum; 
however, unfortunately physical vacuum itself has not been studied much.    
According to the theory we cannot directly measure the {\it physical\/} 
behavior of vacuum particles     
because it is not in real world. Only by using antiparticles, we can guess 
the vacuum particles behavior like a mirror image since we believe 
energy-momentum conservation and charge conservation at least. 
Let us define {\it real world\/} for physical phenomena, {\it imaginary 
world\/} for {\it the intangible one\/} and {\it dual world\/} for both 
the real and imaginary worlds.  
\medskip  
\par  
In stationary vacuum, we can suppose that the electrons, for instance, are not 
packed completely, but that they have spatial gaps among them because we can 
describe a positron using a wave function in quantum mechanics. 
In other words, there must be some uncertainty(${\mit\Delta}{\vec r} > 0$) for 
the position, which means that vacuum electrons also have to have that 
uncertainty in vacuum.   
In addition, vacuum polarization\cite{JDB-2}  
is another clue for the 
spacial gaps.   
For stationary vacuum we can assume that the vacuum particles are 
{\it bounded\/} with negative energies, and that the change of their   
energy produced by any possible perturbation appears in real world because of 
energy conservation. 
Now, if vacuum particles have spacial gaps among them, 
their individual motions can be transfered to neighbors in vacuum. 
Therfore, we can imagine a {\it string vibration\/} for the  vacuum particles, 
which the string consists of.     
In phenomena these vacuum particles(negative energy mass 
particles in vacuum) already have been {\it included\/}, but in measurements 
only real values have meaning in physics.    Even though we concern only real 
values as the final results, we need to care about the interaction  mechanism 
to understand the results with more fundamental way.     
For instance, how can we understand 
electron-positron {\it pair creation\/} and {\it the annihilation\/} without 
vacuum particles?   In fact, the model(Dirac sea) itself is still argumentable among 
theorists, but at least we can accept the existence of vacuum particles 
and the interactions with real physical object.          
\medskip 
\par 
\noindent 
Let us assume followings : 
\begin{itemize} 
\item[(1)]  
  Physical space-time consists of 
  4-dimensional complex space({\it Riemann\/}). 
  That is, each dimension has an internal complex space uniquely, in which 
  any complex function is satisfied with {\it Cauchy-Riemann\/} conditions.   
\item[(2)] 
  In inertial frame, the real time inteval is the same with the imaginary  
  time interval in absolute value. 
\item[(3)]        
  In real world, physical object is in real 3-D subspace with real time 
  even though the object actually has complex coordinates 
  (${\xi}_x, {\xi}_y, {\xi}_z $).   
 \item[(4)]   
  In imaginary 3-D subspace with imaginary time, 
  where vacuum particles reside, physics law is the same as the law 
  in real world. But in measurements, only the change of real values 
  is measured on real coordinates.    
\item[(5)]  
  There are spacial gaps among vacuum particles, and those particles are 
  {\it bounded\/} with negative energy(${E \sim - m_e c^2}$) in the view 
  from real world; however, all possible vacuum states are occupied by each 
  kind of fermions, but let us consider just only stationary vacuum electrons  
  as a model.       
\item[(6)] 
  Any change in imaginary world reflects to real world and vice versa. \\
  For instance, vacuum particles keep interacting with real physical object.   
\item[(7)]  
  Light propagation is made by {\it the vacuum electron-string vibrations\/}   
  in transverse mode. \\   
  The propagation is in the imaginary axis of an internal complex space. 
\end{itemize} 
According to the assumptions above, stationary vacuum electrons are occupied 
in 3-dimensional complex space and keep interacting with real physical 
object in real world.  For instance, how can we understand electromagnetic  
interaction?  Long time ago we explained the interaction with 
{\it action at the distance\/}, and then we became to know that actually 
{\it photons propagate\/} in the interaction.  
Is that reasonable enough to explain the {\it field\/}? How can the photons  
know the direction to propagate? Now we can find a possible explanation  for 
that question:     When a charged or gravitational object is introduced 
to physical space, vacuum particles are rearranged to be more stable; thus, 
we might say this is the {\it physical entity\/} of the field. 
\begin{figure} 
\begin{center} 
\leavevmode 
\hbox{%
\epsfxsize=3.3in 
\epsffile{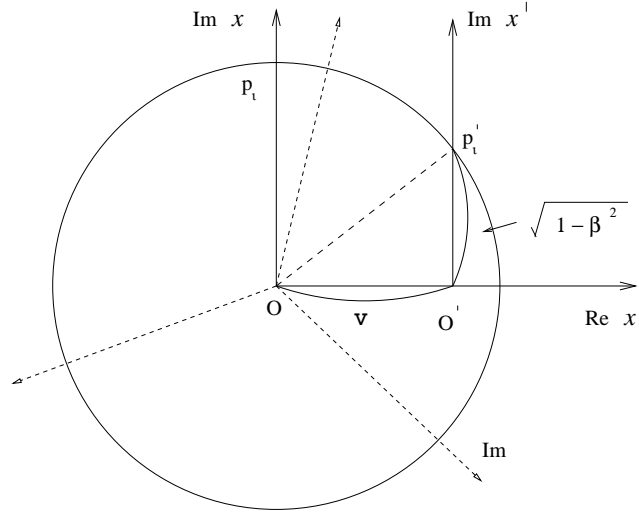}} 
\end{center} 
\caption{3 dimensional complex space($\beta = v/c$)} 
\label{fig2}  
\end{figure} 
\subsection{Time Dilation and Length Cotraction} 
\medskip 
\par 
In Special Theory of Relativity, one of assumptions is about the  
{\it speed of light\/} in free space, which is the same for all inertial 
observers.  Now we need to review this assumption with the physical space 
we are assuming.      
If the propagation of light follows the imaginary axis in an internal complex 
space -- assumption(7), the speed of light is independent of a relativistic 
constant of motion on real axis because of the orthogonality between the real 
and imaginary axis.   
Since the time interval is same on the imaginary and the real axis of 
an inertial frame -- assumption(2),  the scales in both axises are also same, 
the wave front of light in the imaginary axis is corresponded to the real 
axis with  the exactly same coordinate as the imaginary because the change 
of energy in vacuum is mapped to real world -- assumption(6). 
\par \medskip   
In Fig.(\ref{fig2}), there are two observers -- A at the origin and B  
with the velocity($v$) in $x$-direction.  When the moving observer B  
passed through the origin, they sinchronized their watches to zero    
and a light source started radiating from the origin. 
One second later, the observer A at the origin of O-frame recognize 
that the moving observer's time is $\sqrt {1-{\beta^2}}$  
because the light signal in ${O^\prime}$-frame  propagated 
from ${O^{\prime}}$ to ${{p_i}^\prime}$ through the imaginary 
${x^\prime}$ axis and the time interval equals to the imaginary time interval  
in the internal complex space.  However, this time dilation effect is the same 
to observer B since the effect is relativistic.  
\par \medskip
With a similar procedure as above we can also show the length contraction 
effect.  Let us say, in O-frame a rod is appended at the origin and extended
to $x=L_o$ -- proper length.  At $t=L_o/v$, the origin of ${O^\prime}$-frame 
is at $x=L_o$ and $t^\prime = (L_o/v) \sqrt{1-\beta^2}$.  Definitely observer 
A, thus, surmise that observer B estimate the length of the rod as 
$L^{\prime} = L_o \sqrt{1-\beta^2}$.  
\subsection{Internal Complex Space}  
\paragraph{} 
According to the assumptions of physical space,  3-dimensional complex vector 
is 
\begin{equation} 
{\vec \xi} \makebox[1mm]{}=\makebox[1mm]{}\xi_x {\vec i} \makebox[1mm]{}
          +\makebox[1mm]{}\xi_y {\vec j} \makebox[1mm]{}
          +\makebox[1mm]{}\xi_z {\vec k}, \makebox[1cm]{} 
\label{CV} 
\end{equation}  
$$
\makebox[1cm]{}=\makebox[1mm]{}(x, y, z)\makebox[1mm]{}+ 
           \makebox[1mm]{} i (x_i, y_i, z_i). \makebox[1cm]{}  
$$ 
In the expression, ${\vec r} = ( x, y, z )$ and ${\vec r_i}=( x_i, y_i, z_i )$ 
were used for the real and the imaginary coordinates, respectively. 
Then {\it complex length element\/} $d{\vec l}_c$ is defined as  
$$ 
d{\vec \xi} = (dx, dy, dz)\makebox[1mm]{}+\makebox[1mm]{}i (dx_i, dy_i, dz_i), 
$$ 
and the absolute length square is 
$$ 
l^2 \makebox[1mm]{} = \makebox[1mm]{}{\vec\xi} \cdot {\vec\xi^*}, 
\makebox[4cm]{}   
$$ 
$$ 
\makebox[1cm]{} = \makebox[1mm](x^2 +y^2 + z^2) \makebox[1mm]{}
                  +\makebox[1mm]{}({x_i}^2 + {y_i}^2 + {z_i}^2).  
$$ 
Here, the real length, $r = \sqrt{x^2 +y^2 + z^2}$ and the imaginary length,\\ 
$r_i = \sqrt{{x_i}^2 + {y_i}^2 + {z_i}^2}$.  So,   
\begin{equation} 
l \makebox[1mm]{}=\sqrt{ \makebox[1mm]{}r^2 
                    \makebox[1mm]{}+\makebox[1mm]{}{r_i}^2}, \makebox[3cm]{} 
\label{CL} 
\end{equation}  
and the complex length is 
$$
l_c \makebox[1mm]{} = \makebox[1mm]{} r \makebox[1mm]{} 
                     + \makebox[1mm]{} i \makebox[0.5mm]{}  r_i. 
\makebox[3.5cm]{} 
$$ 
Since $\xi_x = x + i x_i$ and ${\xi}^*_x =  x - i x_i$, 
the differential operators are 
$$ 
{\partial \over {\partial \xi_x}} \makebox[1mm]{} = \makebox[1mm]{}   
{1\over 2} ({\partial \over {\partial x}} - 
i {\partial \over {\partial x_i}}), \makebox[2cm]{} 
{\partial \over {\partial \xi^*_x}} \makebox[1mm]{} = \makebox[1mm]{}   
{1\over 2} ({\partial \over {\partial x}} +   
i {\partial \over {\partial x_i}}).  
$$ 
So,
$$ 
df(x,x_i) \makebox[1mm]{} = \makebox[1mm]{} 
      {\partial f \over {\partial x}} \makebox[1mm]{} dx \makebox[1mm]{} + 
\makebox[1mm]{}{\partial f \over {\partial x_i}} \makebox[1mm]{} dx_i,   
\makebox[2cm]{} 
$$ 
$$ 
\makebox[1.5cm]{} = \makebox[1mm]{} 
      {\partial f \over {\partial \xi_x}} \makebox[1mm]{} d\xi_x 
\makebox[1mm]{} + 
\makebox[1mm]{}{\partial f \over {\partial \xi^*_x}} 
\makebox[1mm]{} d\xi^*_x.     
\makebox[2cm]{} 
$$  
In general, 
$$ 
df(x,x_i; y,y_i; z,z_i) \makebox[1mm]{}=\makebox[1mm]{} 
(d\vec x \cdot \nabla \makebox[1mm]{}+
\makebox[1mm]{}d\vec x_i \cdot \nabla_i)\makebox[1mm]{} f. 
$$  
But, in compact way, differential operators can be defined as    
\begin{equation} 
\nabla_l \makebox[1mm]{}\equiv  
\makebox[1mm]{} {\vec i} {\partial \over {\partial l_x}} +    
\makebox[1mm]{} {\vec j} {\partial \over {\partial l_y}} +   
\makebox[1mm]{} {\vec k} {\partial \over {\partial l_z}}, \makebox[1cm]{} 
\label{DEL} 
\end{equation}   
where $l_x$ is the length of complex variable, $x + i x_i$, in the internal 
complex space of $x$-axis. 
That is, $l_x = l_x(x,x_i)$, $l_y = l_y(y,y_i)$ and  $l_z = l_z(z,z_i)$, 
and the $\vec l$ is not unique but dependent on intrinsic variables, that is, 
$\vec l = \vec l(\theta_x, \theta_y, \theta_z)$. 
Hence, there are infinte number of $\vec l$-subspaces made by $\vec l$.    
And Laplacian operator is  
\begin{equation}   
{{\nabla}_l^2} \makebox[1mm]{} \equiv {\vec \nabla_l} 
       \makebox[1mm]{} \cdot \makebox[1mm]{}
       {\vec \nabla_l}. \makebox[3cm]{}     
\label{LAP}  
\end{equation}   
If a physical system is invariant under the rotation in the internal complex 
space -- U(1) symmetry, the Laplacian, ${\nabla_{_l}}^2$, is independent of 
$\theta_x, \theta_y$, and $\theta_z$, and thus equals to  ${\nabla}^2$ 
in {\it 3-dim.\/} real space.             
For a convenience sake, surface element and volume element can be defined as 
\begin{equation} 
{dS}_{xy} \makebox[1mm]{} = \makebox[1mm]{} dl_x \makebox[1mm]{} dl_y,
\makebox[3cm]{} 
\label{CS} 
\end{equation} 
and,
\begin{equation}  
{dV}\makebox[1mm]{} = \makebox[1mm]{} dl_x \makebox[1mm]{} 
                        dl_y  \makebox[1mm]{} dl_z,  
\makebox[3cm]{} 
\label{DV}  
\end{equation} 
in which $dl_x = \sqrt{(dx)^2 + (d{x_i})^2}$. For vacuum particles, mass and 
the mass density are assumed as real.   
\medskip 
\par
\begin{figure} 
\begin{center} 
\leavevmode 
\hbox{%
\epsfxsize=4.4in 
\epsffile{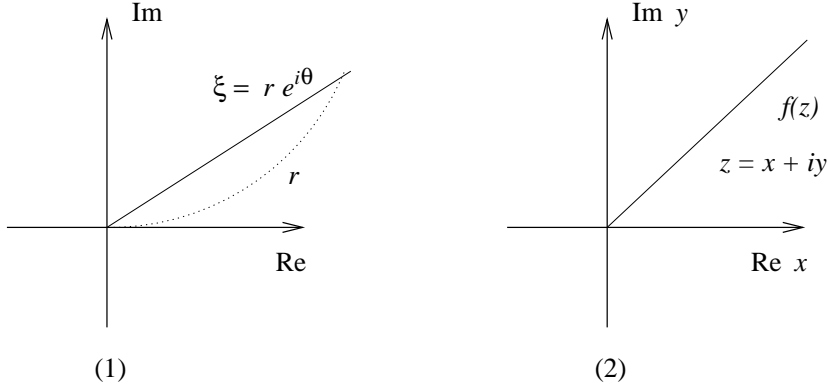}} 
\end{center} 
\caption{internal and coupled complex space} 
\label{fig3}  
\end{figure}

 
Now let us think about what kind of analytic function $f({\xi})$ is 
possible and how the function can be related to a real valued function on 
the real axis.  Fig.(\ref{fig3}-1) shows an internal complex space. 
As an example in electrostatic case, let us say, 
a complex potential function $f(\xi)$ is defined in the internal 
complex space and the function, $f(\xi) = U(x, x_i) + i V(x, x_i)$ 
as an analytic function. 
Then the real function $U(x)$ is unique on real axis  because of 
the {\it uniqueness theorem\/}\cite{UNIQ}. Furthermore, the real function 
$V(x)$ is also unique because of {\it Cauchy-Riemann\/} conditions.
\cite{CAU-1}\cite{CAU-2} That means, function  
$f(\xi)$ is unique on real axis. So the complex potential function $f({\xi})$ 
corresponding to the function $U(x)$ on real axis must be unique because of  
{\it analytic continuations\/}.\cite{ANAL}   
Hence, once we find a complex potential function $f({\xi})$ that is 
corresponded to $U(x)$ on real axis, then the function $f({\xi})$ is 
{\it unique\/}.   
For instance, let us imagine a conducting sheet on ${x = 0}$ plane with 
a uniform surface charge density(${\sigma} > 0$) with assuming that the edge 
effect is ignorable. Then the electrostatic potential,
$$ 
{\phi(x)}\quad = \quad -\makebox[2mm]{}{\alpha}\makebox[2mm]{}{\vert x \vert}. 
\makebox[6cm]{( $\alpha = {2 \pi  \sigma}$ in cgs )}    
$$ 
And the complex potential, 
$$ 
f(\xi) \quad = \quad -\makebox[2mm]{}{\alpha}\makebox[2mm]{}{\xi}.  
\makebox[6.2cm]{}  
$$ 
For another example,  point charge Q is at the origin in three dimensions. 
The electrostatic potential {$\phi(r)$} is isotropic 
and thus the corresponding complex potential is also independent of 
{$\theta $, $\phi $} as following :  
$$ 
\phi(r) \quad = \quad {Q \over r}, \makebox[4cm]{} 
f(\xi) \quad = \quad {Q \over \xi}, \makebox[0.6cm]{}  
$$
with $\xi = r + i r_i$.    
\medskip 
\par   
Using {\it analytic continuations\/}, we can expand the electrostatic 
potential (in  real 3-dimension) to {\it complex potential\/} in  
3-dimensional complex space uniquely. 
And we give the meaning to the complex potential as for complex space, 
in which vacuum particles reside. Without a loss of generality, 
{\it gravitational potential\/} function also can be expanded 
to the corresponding complex potential function.      
\medskip 
\par 
If there is no symmetry, even in two dimensions the complex potential is 
getting complex since we need to manage two real and two imaginary dimensions 
simultaneously.  Also there are couplings  between real {\it x\/} and 
imaginary {\it y\/} or between imaginary {\it x\/} and real {\it y\/} 
in case that the {\it separation variable method\/} is not possible.  
But still we can find the complex potentials in principle as before. 
For instance,  to describe a two dimensional electrostatic case   
let us choose real {\it x\/} and imaginary {\it y\/} --  
coupled complex space --  as shown in Fig.(\ref{fig3}-2).  
The analytic function, 
$$ 
\makebox[2cm]{} f(z) \makebox[2mm]{} = \makebox[2mm]{} U(x,y) \makebox[2mm]{} 
+ \makebox[2mm]{} i V(x,y). \makebox[6cm]{}  
$$ 
Since $U(x,y), V(x,y)$ are harmonic functions, these functions are determined 
uniquely with a boundary condition given in the complex space. The function 
$U(x,y)$ is real 2-dimensional electrostatic potential, that is ${\vec E} = 
- {\nabla U(x,y)}$. In addition, there is a relation between $U(x,y)$ and  
$V(x,y)$ ; ${\vec E} = {- \nabla U(x,y)} = {- \nabla \times {\vec m} V(x,y)}$, 
in which ${\vec m}$ is a unit and normal vector in {\it x-y\/} plane.\cite
{DEN-1}  
\par \medskip
If we choose, instead of the real x and the imaginary y,  the real x and 
real y  -- that is corresponded to $\pi/2$ rotation in the x-internal 
complex space,  
then only the function $U(x,y)$ in the 2 dimensional  real space is necessary 
to get the electric field $\vec E$ and the function $V(x,y)$ disappears.   
With this fact, assumption (4) and (6) urges us to accept U(1) symmetry 
in internal complex space, that is, rotational invariance of phenomenological 
facts --  the absolute length in Eqn.(\ref{CL}) is invariant under the U(1) 
transformation.   Hence, If a potential($ex.$ electrostatic) has a symmetry 
in the real axis, the system has U(1) symmetry --  phenomenological facts -- 
in the entire complex space($-\pi \leq \theta < +\pi$).                   
\subsection{Light Propagation in Vacuum (in free space)} 
\paragraph{}  
At the begining of Section(3), it was assumed that vacuum particles are 
bounded with negative energy in {\it 3-dim.\/} complex space,  and vacuum 
electrons, as a {\it  model\/}, were chosen, each of which has the energy of 
$E \sim - m_e c^2$.   
In stationary vacuum, electrons are supposed to be distributed uniformly and 
spaced equally among them,  and the force is repulsive to each other.   
If there is a small distortion, it is expected that the reaction  seems 
to come from a spring.  Therefore, let us assume 
{\it a simple harmonic oscillation\/}  for a small distortion from 
their equilibrium positions. 
\medskip  
\par 
If light propagation is in imaginary {\it x\/} axis and the oscillation of  
electric field is in  {\it y\/}-internal complex space, then the 
energy propagation is in real {\it x\/} axis and the electric field 
oscillation is in real {\it y\/} axis in the view from real world.  
\begin{figure}    
\begin{center}
\leavevmode
\hbox{%
\epsfysize=1.2in
\epsffile{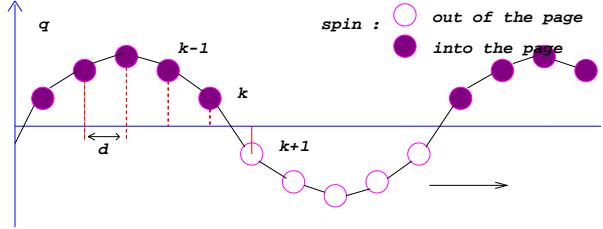}}
\end{center}
\caption{transverse mode.}
\label{fig4}
\end{figure}   
\medskip 
\par
\begin{figure}  
\begin{center}
\leavevmode
\hbox{%
\epsfysize=1.2in
\epsffile{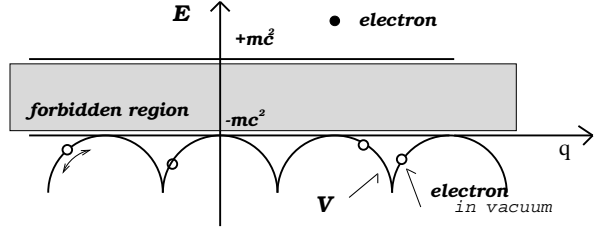}}
\end{center}
\caption{potential energy for small oscillation.}
\label{fig5}
\end{figure}

  
Let us suppose that plane wave propagation is made of one dimensional wave
strings as in fig.(\ref{fig4}) (equilibrium distance d, $\makebox[1mm]{}$ 
spring constance $ {\it K}>0 $) with a potential  
$ V(q) = -{1\over2} {\it K} q^2 $  for each vacuum electron nearby its 
stationary position. In the potential $V(q)$, $q$ represents the image 
on real {\it y\/} axis for the oscillating electron in the internal complex 
space, and the minus sign is used because vacuum electrons have negative 
energy and negative mass. Fig.(\ref{fig5}) represents the potential $V(q)$  
of stationary vacuum electrons for their small oscillations. 
By the way, {\it minimum action principle\/} in real world need to be modified 
for those vacuum particle since they have negative energy and mass.  
\medskip 
\par 
\noindent  
With the lagrangian given as   
\begin{equation}%
L = {1\over 2} \sum_{k=- \infty}^{k=+ \infty} \lbrack (-m_e) \dot {q_k}^2 + 
    {\it K}(q_{k-1} - q_k)^2 \rbrack 
\label{LAG.}
\end{equation}
\noindent 
and using {\it  maximum action principle \/} 
instead of minimum action principle in real world, the equation of motion is  
\begin{equation}%
m_e \ddot q \quad \doteq \quad {\it K} d^2 \quad {\partial^2 q \over 
                                                               \partial x^2}, 
\label{EQM}
\end{equation}
where $d \ll 1$.  
Since tension $S = {\it K} d$ and density $\rho = {m_e \over d}$,  
it leads to a wave equation as    
$$
{\ddot q} \quad = \quad \left({S\over \rho}\right){{\partial^2 q}
                  \over{\partial x^2}}, 
\makebox[1cm]{} 
$$
where $\sqrt{S/\rho}$ is the propagation speed which is equal to the speed of 
light in free space.     
\begin{figure}    
\begin{center}
\leavevmode
\hbox{%
\epsfxsize=4.0in
\epsffile{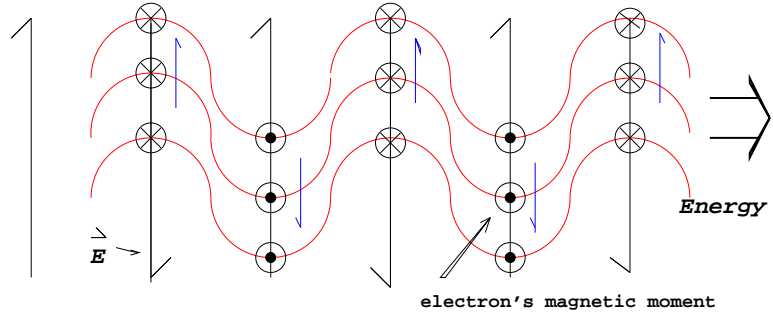}}
\end{center}
\caption{plane wave propagation.}
\label{PROP}
\end{figure}      
\paragraph{}
Fig.(\ref{PROP}) represents the possible mechanism of light propagation 
through vacuum in case that one wave length consists of two vacuum electrons. 
In each energy pulse the electron's motion let only forward spin flip due to
the speed of light.  
\subsection{Interpretation of photon : 
\quad ${E_\gamma} = h \nu = {p_\gamma}c $}
In free space let us suppose that the amplitude in {\it each wave train\/} 
of a plane waves is same for each other and {\it fixed \/} because there is 
no  reason that the amplitudes in plane waves are different.    
If the energy flux is increased in one wave train, the frequency also 
increases because the speed of light is constant or the tension 
$ S \propto \rho $.  Furthermore,  
we know that the energy included in one wave length is proportional to 
the frequency($\nu $)\cite{KEITH}  
since the electron number in one wave length $n_e/\lambda \propto {1\over 
\nu}$  
and the electron's oscillating velocity $ v \propto {\nu}$ in Eqn.(\ref{EQM}).
 What is the proportional constant? 
\par \medskip  
Because we assumed $d \ll 1$ in Eqn.(\ref{EQM}), the discrete wave train is 
presumed as a continuous wave string. 
For the vibrating string (sinusoidal wave), there is a flow of energy down 
the string.   
\noindent
The power\cite{KEITH}  is
$$
{\it P} = k \omega {\it S} {\it A}^2 \sin^2(k x - \omega t),
\makebox[2mm]{} 
$$
where $A$ is amplitude.  Thus, the time average is given as 
\begin{equation}%
\langle {\it P} \rangle_{av} = {1\over 2} k \omega {\it S}{\it A}^2. 
\makebox[2cm]{} 
\label{AVP}
\end{equation}
\noindent 
From Eqn.(\ref{AVP}), the energy included in one wave length is  
$$
{\it E}_{_{\lambda}} = \langle {\it P} \rangle_{av} \makebox[1mm]{} 
{1\over \nu} = 2{\pi}^2 {\it S} {\it A}^2 \makebox[1mm]{} {1\over \lambda}.   
\makebox[2mm]{}
$$
\noindent
On the other hand, ${\it E \/} = p c $  in Special Theory of Relativity.  
Hence, for one wave length the energy ${\it E\/}_\lambda$ is 
$$ 
p_{_{\lambda}} c = 2{\pi}^2 {\it S} {\it A}^2 \makebox[1mm]{}{1\over \lambda}.
\makebox[2cm]{} 
$$ 
The momentum included in one wavelength is        
\begin{equation}   
p_{_{\lambda}} = {1\over \lambda}\makebox[2mm]{}\left[2 {\pi}^2 c {\it A}^2 
   \left({m_e \over d}\right)\right], 
\label{PHOTON}  
\end{equation} 
in which we used relations, $S = \rho c^2$ and $\rho = {m_e \over d}$.  
\noindent
In Section(2), we confirmed : there is a unversal constant for   
light only if we can assume that the amplitudes of the light wave are 
{\it same\/}.  
In other words, if we can assume that there is a basic unit in the amplitude 
of light, then the constant for one wave length in the wave strings, 
by which light propagation is made, is minimum and unique.  
Hence, the universal constant can be expressed as following     
\begin{equation}%
\makebox[3.0cm]{}  
{p_{_{\lambda}}} \lambda = 2 {\pi}^2 m_e c \left({{\it A}^2 \over d }\right).
\makebox[4cm]{}  
\label{PLANK}
\end{equation}
\noindent
As we expected, it is a constant indeed as far as the amplitude {\it A\/} 
is fixed. (Here, $A$ is the amplitude for each wave string.)    
Considering that the energy included in one wave length can be an 
energy flux unit in a wave string, it is natural to accept  this constant 
as Planck's constant({$ h $}).  
From this calculation we come to know that the amplitude {\it A\/} in each 
wave string is {\it constant\/}  and directly proportional to $\sqrt{d}$.
 In addition, this relation is independent of the speed of light.

\section{physical momentum and wave function in QM.} 
\paragraph{} 
The inertial mass in Newtonian physics is no more constant in Special Theory 
of Relativity because energy-momentum-four-vector conservation is assumed 
in {\it Minkowski space\/}(real world). But the theory says, the mass increase 
-- energy gain -- is independent of interactions between two inertial frames 
or measurements. how is that possible?  
\medskip \par 
If mass  $M$ is at the origin of $O^\prime$-frame in Fig.(\ref{fig1}) and 
$O^\prime$-frame is moving away from $O$-frame with a constant velocity, 
$v =\beta c$, then the mass in $O$-frame is $M \gamma$  (${\gamma}^{-1} = 
{\sqrt {1 - \beta^2}}$).  Since vacuum particles are interacting with 
physical object -- assumption(6), it is necessary to check out how 
the distribution of vacuum particles has been changed.  However,  
gravitational interaction between mass $M$ and vacuum particles is repulsive, 
and thus the potential energy is positive  because vacuum particles have 
negative masses.   In the mass rest frame, 
$O^\prime$-frame, the distribution is isotropic in 3-dimensional complex 
space, but in $O$-frame, it is not isotropic anymore because of the 
{\it length contraction\/} effect. In $O$-frame, 
the length contraction effect results in the mass density($\rho$) increase 
(in absolute value) with the factor of $\gamma$  
because total mass of vacuum particles has not been changed, but the volume of 
space has been changed with the factor of $\gamma^{-1}$.  
  
\medskip 
\par 
Firstly, in the rest frame ($O^\prime$-frame) ; let us say,  
the potential energy of the mass $M$ in the interactions with vacuum particles 
is $|\mit\Phi_0|$ -- $U(1)$ symmetry.  
Then, in $O$-frame ; the potential energy  
$|\mit\Phi| = \gamma \makebox[1mm]{} |{\mit\Phi}_0|$  
because the mass density in vacuum has been increased with the factor 
of $\gamma$  --  the mass distribution of vacuum particles is assumed as 
continuous ; the space of vacuum, as infinite. 
If we compare gravitational potential energies ;   
in $O^\prime$-frame, 
$$ 
|{\mit\Phi_0}| \makebox[1mm]{} = \makebox[1mm]{}  
\int_{\mbox{\it vac.\/}} G M 
\makebox[1mm]{}{\rho(\xi)\over{|\xi|}}\makebox[5mm]{} d{\vec l}, 
\makebox[0.5cm]{} 
$$
and in $O$-frame, 
\begin{eqnarray}
|\mit\Phi|  &=& \int_{\mbox{\it vac.\/}}\gamma\makebox[1mm]{} G M 
          \makebox[1mm]{}{\rho(\xi)\over {|\xi|}} \makebox[5mm]{} 
                 d{\vec l},                            \nonumber \\
            &=&  \gamma\makebox[1mm]{}  |\mit\Phi_0 |,  \nonumber 
\end{eqnarray}
in which $\xi = r + i r_i$, $d{\vec l} = dl_x dl_y dl_z$ and $\rho(\xi)$ is 
the mass density of vacuum particles in absolute value ;      
then,  it is natural to acknowledge the inertial mass increase 
in the relativistic motion.  But, in fact, it is not true  because the mass 
itself is not increasing but has the effect from the vacuum. 
\medskip 
\par 
In Section(3), vacuum electron-strings vibrations, following imaginary axis, 
was considered as light wave in real world, and  
a basic unit in the amplitude was assumed.  Then, it came up 
that the unit of momentum $p$ of light  is $\hbar k$ for each $k$.   
To investigate {\it physical momentum\/} which is possibly related with 
vacuum particles, let us say, the velocity $v$ is in $x$ direction 
in Fig.(\ref{fig1}) as a simple case.  Then, in $O$-frame; the effect of 
interactions with vacuum particles is moving with the mass $M$  and   
the interaction is  symmetry about $x$ axis.  
Thus, we can bring ({\it mapping\/}) all interaction effects into the internal 
complex space.    
In $O$-frame, the energy formula is  
\begin{equation}   
E = \sqrt{M^2 c^4 + p^2 c^2}, \makebox[2cm]{}  
\label{ENEG} 
\end{equation}
$$ 
\makebox[4cm]{}=\makebox[1mm]{}M\gamma c^2, 
\makebox[2cm]{} (\gamma ={1\over{\sqrt{1 - \beta^2}}}) 
\makebox[2.4cm]{} 
$$   
and the momentum, $p = M \gamma  \beta c$.  In Eqn.(\ref{ENEG}), the second 
term in the square root 
must be come from the interaction between the mass $M$ and vacuum particles. 
In real world, if the interaction effects with vacuum particles 
is not considered,  only the mass $M$ has {\it all\/} physical quanties, 
such as energy, momentum, etc. 
But it needs to be reconsidered in fact that the interaction is included in 
the system.
\begin{figure} 
\begin{center} 
\leavevmode 
\hbox{%
\epsfysize=2.2in 
\epsffile{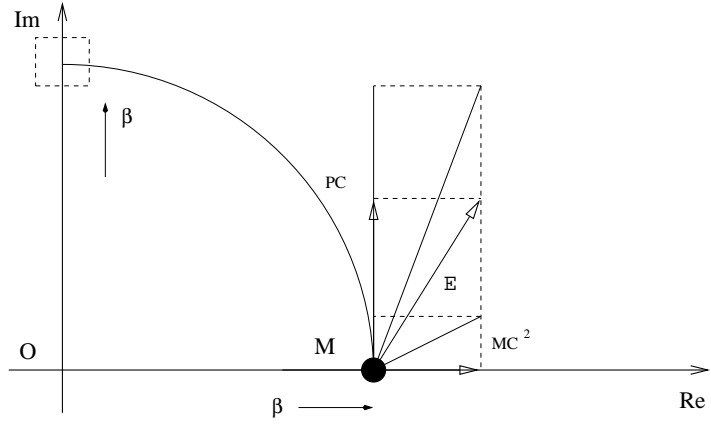}} 
\end{center} 
\caption{internal complex space with mass $M$ in $O$-frame} 
\label{KT}  
\end{figure}         
\noindent  
Special Relativity says, if the mass $M$ goes to zero, the energy and momentum  is 
also zero. But light still has momentum even though the mass is zero; however, 
that is a singularity in Eqn.(\ref{ENEG}) and the equation is not definite 
when the mass goes to zero with $\beta$ goes to 1.           
In another words, the momentum is dependent on mass if the mass is not zero, 
but the momentum is also independent of the mass in light case.  
\medskip 
\par    
To be consistent in the interpretation of the energy formula 
in Eqn.(\ref{ENEG}),  
the second term, $p c$,  should be the energy carried by vacuum particles 
in a form of bundle of pulse signals  following the physical object : 
Hence, {\it physical momentum\/} in real world is the representation 
of the interaction with vacuum particles, and in fact vacuum particles 
carry the momentum on the imaginary axis of the internal complex space.     
Now there is another question why these two terms are not direct sum in the 
energy formula. That is because the second term, $p c$, came from 
the interaction between the mass $M$ and vacuum particles.  
In Section 3.3, we introduced U(1) symmetry in internal complex space ; 
that means the phenomenological facts -- energy in this case -- is invariant   
under the unitary transformation, U(1), as shown in Fig.(\ref{KT}).       
The rest mass energy  $M c^2$ is not changed in a relativistic motion and 
thus independent of the second term $p c$, which is on the imaginary axis 
since the energy comes from the interaction with vacuum particles and 
the interaction follows the imaginary axis.  
Meanwhile,  there is an interesting point : 
In magnitude, $p c$ is always greater than the kinetic energy in $O$-frame as 
far as both are not zero.  
That is because vacuum particles have negative mass and  thus they need some 
energy when they back to equilibrium states. 
\medskip  
\par
If a physical object is in one dimensional motion as above,  the effects of 
the {\it interaction with vacuum particles\/}  can be expressed with 
one dimensional wave function in real world.  Let us say, a complex function  
${\cal F}(x,t)$ represents the effects, which is following the physical object 
in a wave form. 
Hence, the wave function ${\cal F}(x,t)$ represents not a phenomenal wave but 
the result of summation of infinte number of wave strings in vacuum, and  
the total energy  is $P c$  which the wave is carrying.       
The wave function, {\it in general\/}, can be given as 
\begin{equation}    
{\cal F}(x,t) \makebox[2mm]{} = \makebox[2mm]{} 
{1\over \sqrt{2\pi}}\int_{-\infty}^{\infty}\makebox[2mm]{}{\cal A}(k)
\makebox[2mm]{}e^{{i}[kx - \omega(k) t]}\makebox[2mm]{} dk.    
\label{FS} 
\end{equation} 
Here, ${\cal A}(k)$ is the amplitude of wave string which has wave vector  
$k$, and $\omega(k)$ is angular velocity. 
Eqn.(\ref{FS}) is a general expression as long as we can assume 
{\it superposition principle\/} ; however, there is a condition on 
$\omega(k)$  because the propagation is in the positive $x$-direction.  
The condition is 
\newpage 
\begin{eqnarray}
\omega(k) &\geq& \makebox[1.5cm]{for} k \geq 0,  \nonumber \\
          &\leq& \makebox[1.5cm]{for} k \leq 0,  \nonumber
\end{eqnarray} 
and continuous.  Also we can assume $A(0)=0$ and $\omega(0)=0$ without a loss 
of generality.    
Since the vacuum is not homogeneous as in  plane waves,  
$\omega(k) \over k$ is not a constant but can be even greater than $c$,   
speed of light in free space.   
\medskip 
\par  
Even though the function ${\cal F}(x,t)$ represents the effects from vacuum 
particles, only real value({\it ex.\/} energy) has meaning in real world. 
Now we can find the necessary conditions of ${\cal F}(x,t)$  using a  
formula, $-{1\over 2} S({{\partial f} \over {\partial x}})
          ({{\partial f}^* \over {\partial t}})$  that is average power 
for a wave,  $f(x,t) = A e^{i(kx -\omega t)}$, where $A$ is amplitude. 
Let us say, the wave is extended from $x=-L$ to $x=L$({\it e.g.\/} 
$L\rightarrow \infty$ later), then the average power of the wave is 
$$   
<Power>\makebox[1mm]{}=\makebox[1mm]{} -{1 \over 2}\cdot{S \over 2 L}
                      \makebox[1mm]{}
          \int_{-L}^L dx \left({\partial {\cal F}\over {\partial x}}\right) 
                       \left({\partial {\cal F}^* \over {\partial t}}\right),  
$$
in which we used the orthogonality of trigonometric functions.   
Since the average power is the average energy density  times group velocity
($v_g$),  total energy carried by the wave is 
\begin{eqnarray} 
P c  &=& M \gamma v_g c  \nonumber \\ 
     &=& {1\over v_g} {S \over 2} \int_{-\infty}^\infty dk \makebox[1mm]{} 
                   |{\cal A}(k)|^2 k \makebox[1mm]{} \omega(k). 
\label{FIN}  
\end{eqnarray} 
In which $P$ is the momentum of the particle, $\delta$-function was used, and 
$L$ went to infinity.  In Eqn.(\ref{FIN}) the integration on RHS must be 
finite as far as the energy on LHS is not infinity.  Hence, the integration, 
$\int_{-\infty}^{\infty} dk |{\cal A}(k)|^2$ must be finite 
because $k\cdot\omega(k) > 0$.   That is {\it square integrable\/}. 
Now if we can assume that the function, ${\cal A}(k)$ is continuous 
in the region( $-\infty< k <\infty$),   
then we can use {\it Fourier transformation\/}.   
Hence, the function, ${\cal F}(x)$ is also square integrable because of 
{\it parseval theorem\/}\cite{PARS}.   
\medskip 
\par 
In general, Using Fourier transformation, the function  ${\cal F}(x,t)$ and 
the conjugate function ${\cal G}(k,t)$ in $k$-space with $\omega(k)$ 
are given as  
$$  
{\cal G}(k,t) \makebox[2mm]{}=\makebox[2mm]{}{1\over{\sqrt{2 \pi}}} 
\int_{-\infty}^{\infty}\makebox[2mm]{}{\cal F}(x,t)
\makebox[2mm]{}e^{-{i}[kx - \omega(k) t]}\makebox[2mm]{} dx,    
$$ 
and   
\begin{equation}  
{\cal F}(x,t) \makebox[2mm]{} = \makebox[2mm]{}{1\over{\sqrt{2 \pi}}} 
\int_{-\infty}^{\infty}\makebox[2mm]{}{\cal G}(k)
\makebox[2mm]{}e^{{i}[kx - \omega(k) t]}\makebox[2mm]{} dk,   
\makebox[4mm]{} 
\label{FXT} 
\end{equation} 
where ${\cal G}(k,t) = {\cal G}(k) e^{-i \omega(k) t}$.  
In Eqn.(\ref{FIN}), $|{\cal A}(k)|^2$ can be interpreted as it is proportional 
to the energy density in $k$-space in comparison with Eqn.(\ref{AVP}), 
then $|{\cal F}(x)|^2$ is also proportional to the energy density 
because of the symmetry property in Fourier transformation. 
Therefore, it is natural to interpret that $|{\cal F}(x)|^2$ is proportional 
to {\it probability density\/} in single particle case.   
\medskip 
\par 
In Eqn.(\ref{PLANK}) $p \lambda$(Plank's constant, $h$)   
is invariant under Lorentz transformation and General Relativity says, we can 
always construct MCRF(momentary co-moving reference frame), that is, 
a momentary Lorentz inertial frame.  Therefore, we can assume that Plank's 
constant($h$) is invariant even under non-uniform space(interaction) 
because Lorentz transformation is a {\it continuous\/} transformation.    
With relations for one photon; $p = \hbar k$ and ${\epsilon\/} = \hbar \omega$ 
(but $ {\epsilon\/} \not = p c$.), 
Eqn.(\ref{FXT}) is  
\begin{equation}    
\psi(x,t) \makebox[2mm]{} = \makebox[2mm]{}{1\over{\sqrt{2 \pi \hbar}}} 
\int_{-\infty}^{\infty}\makebox[2mm]{}\widetilde{\psi}(P)
\makebox[2mm]{}e^{{i \over \hbar}[P x - E(P) t]}\makebox[2mm]{} dP.    
\label{QMF} 
\end{equation} 
Here, $P$ and $E(P)$ represent the energy and momentum of the particle, 
respectively,  and a scale transformation was used.  
In the scale transformation,  we used a fact that the system energy is finite. 
 Now, function ${\psi(x,t)}$ is nothing but the wave function in 
Quantum Mechanics.  Furthermore, the uncertainty relations such as  
{\small{$\Delta$}}$E$ {\small{$\Delta$}}$t \geq$ {\large O} and   
{\small{$\Delta$}}$x$ {\small{$\Delta$}}$p \geq$ {\large O} 
({\large O}$ \sim \hbar/2, \hbar, h, \cdots$)   can be derived from 
Eqn.(\ref{QMF}) if we know $\psi(x,t)$ or $\widetilde{\psi}(P)$ -- 
as a statistical criterion  ; moreover, 
the relation is consistent to the length  contraction effect in 
Special Relativity -- in phenomenological facts.  

\def\DELTA{$\Delta$}
\section{Interpretation and Experiments} 
\subsection{Wave Function Collapse} 
According to our model interpretation, the wave function in quantum mechanics 
is not the physical object itself but the representation of the interactions 
between the physical object and vacuum particles, in which the physical object 
cannot be isolated without disturbing the wave packet -- 
interaction pattern({\it symbolic\/}) of the vacuum particles.  Therefore, 
the uncertainty  principle is indispensable to describe phenomenological 
facts. In other words, nature itself -- in phenomena -- 
is not apt to the {\it determinism\/} and  {\it locality\/}.  
Thus, we can interpret quantum mechanical formalism as 
a phenomenological realization of nature in fact that mother nature herself 
has a statistical property.   
\par \medskip 
Considering that the question of Wave Function Collapse was originated from 
the view as : the wave function itself is the physical object,  and  
the detection is determined without any time delayed, we don't need to 
consider the question because the interacting time({\small{\DELTA}}$t$) 
with a detector cannot be zero -- wave function never be collapsed --  and 
we are dealing with phenomenological statistics -- though it has been the best 
way up to now.  
\subsection{One particle double slits experiment} 
\begin{figure}  
\begin{center}
\leavevmode
\hbox{%
\epsfxsize=16cm
\epsffile{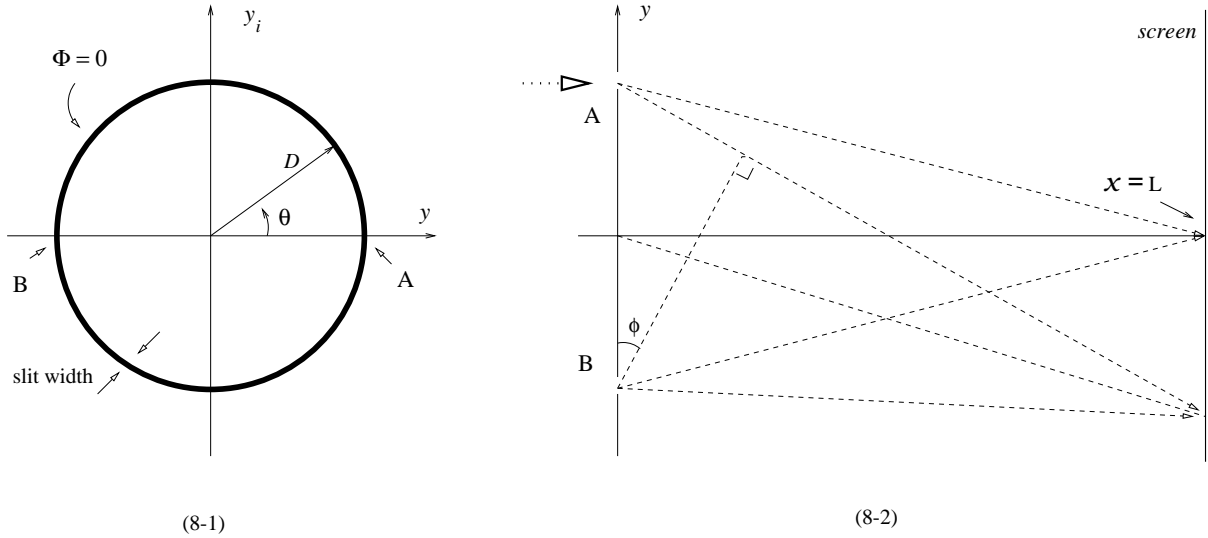}}
\end{center}
\caption{wave function interference }
\label{fig6}
\end{figure}

As shown in Fig.(\ref{fig6}-2), the slits are at $y=\pm D$ and the screen is 
at $x=L$.  If one particle({\it ex.\/} neutron) is incident at $y=D$(slit A), 
the perturbation on stabilized vacuum also affects on slit B simultaneously  
because of $U(1)$ symmetry and analytic condition  
as shown in Fig.(\ref{fig6}-1).   
Now, let us say,  effects of the perturbation can be given in y-internal 
complex space as  
\begin{equation} 
\phi(x,D,\theta,t)\makebox[1mm]{}=\makebox[1mm]{} \int_{-\infty}^{\infty}  
       \left[ \sum_{n=-\infty}^{\infty} A_n(k,D) e^{i n \theta}\right] 
       e^{i(kx - \omega(k) t)} dk,  
\label{SLITS} 
\end{equation}
in which we assume that the slit width is ignorable and the source is 
far away from the slits.   In Eqn.(\ref{SLITS})  $A_{+n} = A_{-n}$ 
since results of the interference must be on real y-axis.             
Now, instead of chasing the particle itself we can choose an anternative way 
-- wave function interference --  to locate the particle position on the 
screen ; because the particle cannot be isolated from the wave packet, and 
the wave function, which is presenting the wave packet, is related to 
the particle's position with probability. 
\par \medskip 
If the incident particle is in a relativistic motion($\gamma \gg 1$), then 
the particle is highly localized with {\small{\DELTA}}$x \ll$ {\large O}  
but {\small{\DELTA}}$p \gg$ {\large O}.  
Moreover, duration time of the perturbation 
on slits is almost instantaneous({\small{\DELTA}}$t \ll$ {\large O}) ; hence, 
the possible interference pattern can be observed only at around center of 
the screen($x=L$, $y=0$).     
To get a simple picture of the interference, let us say, 
{\large O} $ \sim \hbar$ and the incident particle is in the relativistic 
motion($\gamma \gg 1$); thus, the particle is highly localized as 
{\small{\DELTA}}$x \ll \hbar$ but {\small{\DELTA}}$p \gg \hbar$.  Then, 
$\phi_{max}$ in Fig.(\ref{fig6}-2) can be estimated using the uncertainty 
relation({\small{\DELTA}}$E$ {\small{\DELTA}}$t \sim \hbar$) : 
the maximum path difference, 
\begin{eqnarray} 
\makebox[2cm]{} 2D sin \phi_{max} &\doteq& v_g (\Delta t), \nonumber \\ 
                  &\sim &  {v_g \hbar \over \Delta E} \makebox[1cm]{} 
               \makebox[3cm]{($\Delta E$ $\Delta t \sim \hbar$)} \nonumber \\  
                  &\sim &  {\hbar c\over \Delta E} \makebox[1cm]{} 
               \makebox[3cm]{($E \sim pc $)} \nonumber
\end{eqnarray}
If $\gamma \gg 1$, then $sin \phi \sim \phi$. Thus,  
$$ 
\phi_{max} \sim {\hbar c\over 2 D (\Delta E)}. 
$$ 
Using this fact we can further estimate uncertainty of the particle position 
on $y$-axis({\small{\DELTA}}$y$).  That is 
$$ 
\Delta y \sim \pm {L \hbar c \over 2 D (\Delta E)}.  
$$ 
Here, $\Delta E$ is the energy uncertainty of the particle, 
and $\pm$ sign was used because 
we cannot distinguish which slit the particle has passed through; hence,    
the position distribution of particles on the screen should be symmetry.

\subsection{photoelectric effect}
photoelectric effect is called for the interaction between one photon and 
a bounded electron in a metal.      
As long as we can assume the interaction is individual -- 
one vacuum-particle-string oscillation and one bounded electron,  we can also 
confirm that the interaction is almost instantaneous
({\small{\DELTA}}$t \sim \lambda / c$) : 
According to our intepretation of photon in  Eqn.(\ref{PHOTON})(\ref{PLANK}), 
the energy and momentum of one photon is corresponded to the energy and 
momentum carried by one wave length in vacuum-particle-string oscillation. 
Therefore, the maximum kinematic energy which the bounded electron can receive 
is $\sim$ one photon energy, and the interaction time, 
{\small{\DELTA}}$t \sim \lambda / c = 1/\nu$.      

\section{Conclusion} 
\paragraph{} 
The physical entity of wave function in Quantum Mechanics was searched with 
a model suggested in $\S$(3.1) as following :  
\medskip 
\par
Like two sides of a coin - head and tail -- we introduced 4-dimensional 
complex space for physical space and further we defined  U(1) symmetry  
in internal complex space because absolute length is invariant 
under the rotation in internal complex space.  Then, we showed  why light 
speed is constant for all inertial observers and how {\it time dilation\/} 
effect and {\it length contraction\/} effect in Special Theory of Relativity  
can be explained; moreover, we interpreted the energy-momentum formula
(Eqn.(\ref{ENEG})). 
\medskip 
\par       
When the physical object has kinetic energy, it is wrapped with vacuum 
particles(vacuum electrons in the model) those of which are interacting 
with the physical object; the effect of interaction is following 
the physical object in a wave form,  and the velocity($v_g$) is the same as  
that the physical object has.   
In addition, those vacuum particles, {\it in fact\/}, carry the physical 
momentum $p$ -- momentum of the object conventionally -- and kinetic energy  
$p c$.  
\medskip 
\par     
In phenomena, we cannot single out the physical object without disturbing 
vacuum particles, those of which are  wrapping the physical object. 
Therefore,  it is indispensable to accept  {\it uncertainty principle\/} and 
{\it de Broglie matter wave\/} in Quantum Mechanical formalism to describe 
phenomena.   
 In other words, nature itself  has a statistical property intrinsically, 
which means {\it indeterminism\/} and  {\it non-locality\/} in phenomena.     
Thus, the wave function in Quantum Mechanics is not the physical object itself 
but an alternative descritpion -- especially in microphysics --  
for vacuum particles interacting with the physical object.       
\par \medskip
In  $\S$(5), fundamental questions -- Wave Function Collapse, 
one-particle-double-slits experiment -- mentioned in the introduction, and 
photoelectric effect were reviewed.  Especially, one-particle-double-slits 
experiment was explained, and a possible interference pattern was also 
estimated.

\newpage
\paragraph{}
\noindent 
{\bf Figure Caption\/} 
\par 
\medskip  
\begin{itemize} 
\item[Fig.(1)] lorentz transformation with arbitrary $\vec \beta$. 
\item[Fig.(2)] 3-dimensional complex space($\beta = v/c$). 
\item[Fig.(3)] internal and coupled complex space. 
\item[Fig.(4)] light propagation(transverse mode). 
\item[Fig.(5)] potential energy of vacuum electrons for small oscillation. 
\item[Fig.(6)] plane wave propagation. 
\item[Fig.(7)] internal complex space with mass $M$ in $O-$frame.  
\item[Fig.(8)] wave function interference.  
\end{itemize}
%
\end{document}